\documentclass[prb,aps,epsf]{revtex}
\begin{document}
\draft
\twocolumn[\hsize\textwidth\columnwidth\hsize\csname @twocolumnfalse\endcsname

\title{Ratchet-Induced Segregation and Transport of Non-Spherical Grains}
\author{J.F.~Wambaugh$^{1}$, 
C.~Reichhardt$^{1,2}$, and
C.J.~Olson$^{2}$\footnotemark[1], }
\address{$1. \ $ Center for Nonlinear Studies, 
Los Alamos National Laboratory,
Los Alamos, NM 87545 \\
$2. \ $ Theoretical and Applied Physics Divisions, Los Alamos National Laboratory, Los Alamos, NM 87545}

\date{\today}
\maketitle
\begin{abstract}
We consider through simulations the behavior of elongated grains on a vibrating
ratchet-shaped base.  We observe differences in layer velocity profile and
in net grain velocity for grains that are composed of one, two, or three
colinear spheres.  
In the case of mixtures of different species
of grains, we demonstrate layer-by-layer variation in the average velocity
as well as layer segregation of 
species, and show that horizontal separation of the species can
be achieved using this geometry.  
We also find that the addition of a small number of shorter
grains to a sample of long grains provides a lubrication effect that
increases the velocity of the long grains.
\end{abstract}
\vskip2pc]
\narrowtext

\footnotetext[1]{corresponding author : olson@t12.lanl.gov}

\section{Introduction}

Granular systems display a rich variety of phenomena that
are often counterintuitive and in many cases not yet fully understood
\cite{intro,JaegerRMP}.
A notable example is the size-segregation effect,
observable in the Brazil nut problem and related systems.
Mixtures of dissimilarly-sized particles can 
separate by grain
when set in motion by a number of methods, including vertical shaking 
\cite{JaegerRMP,vshake}, horizontal
shaking \cite{Mullin00}, white noise driving \cite{Hong01},
rotating drums \cite{rdrum}, 
rotating boxes \cite{Awazu00},
and shear flow \cite{shear}.  
The segregation, which may be caused
by convective grain flow or by local grain rearrangements under gravity,
can lead to a stratification of mixtures into nearly 
homogeneous layers.

Granular media also display a wide range of behavior when vertically vibrated 
upon an asymmetric sawtooth-shaped base \cite{Farkas99}.
Lining the base with asymmetric teeth breaks the symmetry of the AC 
driving force, leading to net horizontal motion from symmetrical, vertical 
drives.  Such rectification of a fluctuating force leads to a 
ratcheting effect observed in many other systems \cite{genratchet}.
Both experiments and simulations of granular ratchets show that the response 
of a particular type of grain to a ratchet system depends significantly upon 
the size and shape of the ratchet, the driving force, and the properties of 
the grains themselves 
\cite{Farkas99,simulation1,Rapaport01,simulation2a}.
Whether a particular combination of grain and ratchet will display net motion 
to the left, to the right, or at all is not apparent by simple inspection, 
highlighting the need for further study of these systems.
Recently, it was shown that this behavior depends upon how the velocity of 
grains on a ratchet varies vertically, and that specific types of grains 
interact differently with specific base geometries \cite{simulation1}.

Very recent experiments \cite{Derenyi98,Farkas99} and 
simulations \cite{simulation1,simulation2a} have demonstrated that 
the direction in which spherical \cite{footnote1} grains 
move on the ratchet depends on the size and elasticity of the grains.
Thus two species of grains may move in opposite directions on the same
ratchet base.
Since the direction of the velocities persists when the different particles
are mixed, this effect has been used to segregate mixtures of 
particles, allowing for the construction of novel granular sieves that may 
be applicable to practical problems insoluble with 
more conventional,
filter-type sieves.

Although granular ratchets have been successfully modeled in the past using 
spherical grains, the transport behavior of nonspherical grains on 
ratchets has not been considered.
Since there are many examples of extended or nonspherical grains in nature, 
understanding their behavior is of great importance, and has attracted
considerable attention \cite{nonsphere1,nonsphere2,nonsphere3,extendedgrains1}.
In previous studies of anisotropic grains, it was found that the 
static and dynamic responses of the grains depend strongly upon the degree
of anisotropy of the grain \cite{extendedgrains1,extendedgrains2}.
Thus it is of interest to consider the effect of grain elongation on the
behavior of grains in a ratchet system.

In this paper we study the average velocity of elongated grains moving on 
shaken ratchet-shaped substrates.
For extended grains we find that the rotation of the grains plays an
important role in their 
transport.  
We observe both a variation of average grain velocity with depth, and
vibration-induced size-dependent 
stratification of grains.
We demonstrate that it is possible to segregate mixtures of extended grains of 
different lengths by creating conditions where the average velocity of 
different monolayers of grains is in opposite directions for layers 
composed of different grains.
Finally, we show that shorter grains can provide a lubricating effect for the 
motion of longer grains in mixtures.

\section{Simulation}

We consider extended grains 
composed of two or more spheres that share a common axis and are confined to
two dimensions, as in recent experiments \cite{2Dexperiment}.
Each sphere that composes an extended grain is a constrained 
monomer grain, which we refer to as a ``sub-grain''.
Though it is possible to simulate larger grains with our model, we focus
on extended grains composed of one, two or three sub-grains, 
referred to hereafter as monomers, dimers and trimers.
The sub-grains are simulated according to a model for monomer spheres
\cite{Gallas92L,Luding94}, with the addition of a constraint to 
maintain the sub-grains of a single grain fixed with respect to each other.
The sub-grains composing one extended grain are separated by a distance 
equal to the relaxed position of the elastic force governing their 
interaction.
This introduces a rotational degree of freedom 
and corresponding moment of inertia 
effects.
Though three dimensional studies have been conducted in the past 
\cite{simulation1}, ratchet behavior is essentially a two-dimensional 
effect and the complexity inherent to additional dimensions is not 
considered here.

The equation of motion for the monomers or sub-grains is \cite{MD}:

\[m_{i}{\bf \dot{v}}_{i}
= {\bf f}_{\rm el}^{(i)} + 
{\bf f}_{\rm diss}^{(i)} + {\bf f}_{\rm shear}^{(i)} + 
{\bf f}_{g} + {\bf f}_{\rm fric}\]

where the gravitational force ${\bf f}_{\it g} = -0.05$ and 
${\bf f}_{\rm fric} = 0.25$ is a dissipative frictional force corresponding 
to the drag against the confining walls in 2D experimental setups 
\cite{extendedgrains2,2Dexperiment}.
Two grains interact when they are separated by a distance smaller than twice 
their effective radii, $r_{g} = 0.4$.
In this case, the restoration force is given by
\[{\bf f}_{\rm el}^{(i)} = \sum_{i \neq j}{k_{g}m_{i}}\frac{(|{\bf r}_{ij}| - r_{g}){\bf r}_{ij}}{|{\bf r}_{ij}|},\]
where $k_{g} = 20$ is the strength of the restoration spring constant, 
$m_{i} = 1$ is the mass of a sub-grain and 
${\bf r}_{ij} = {\bf r}_{i} - {\bf r}_{j}$ is the vector 
between two sub-grains at ${\bf r}_{i}$ and ${\bf r}_{j}$.
The dissipation force due to grain elasticity  is 
\[{\bf f}_{\rm diss}^{(i)} = -\sum_{i \neq j}\gamma m_{i}
\frac{({\bf v}_{ij}{\bf \dot{r}}_{ij}){\bf r}_{ij}}{|{\bf r}_{ij}|^{2}},\]
where $\gamma = 0.48$ is a phenomenological dissipation coefficient and 
${\bf v}_{ij} = {\bf v}_{i} - {\bf v}_{j}$ is the relative velocity.
The shear friction force has the form
\[{\bf f}_{\rm shear} = - \gamma_{s} m_{i}\frac{({\bf v}_{ij}{\bf \dot{t}}_{ij}){\bf t}_{ij}}{|{\bf r}_{ij}|^{2}},\]
where $\gamma_{s} = 1.2$ is the coefficient of shear friction and 
${\bf t}_{ij} = (-r^{y}_{ij},r^{x}_{ij})$ is the vector ${\bf r}_{ij}$ 
rotated by $90^{\circ}$.  
The parameters were chosen as a compromise between producing realistic 
results and relatively short simulation time.  
The behavior we observed is not qualitatively changed by different parameter 
choices.

Though molecular dynamics (MD) 
grains are soft in the sense that they interact via spring forces 
instead of absolutely defined radii, an effective radii 
$r_{\rm eff} = 0.2$ of 

\begin{figure}
 \centerline{
 \epsfxsize=3.3in
 \epsfbox{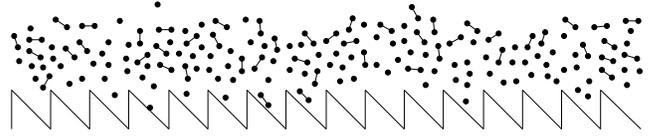}}
\caption{
Snapshot from simulation of a mixture of $N_{1} = 100$ monomer and $N_{2} = 50$ dimer grains.
Sub-grains and monomers appear as filled circles and the stiff bond between the dimer sub-grains is drawn as a line.
After being dropped individually from above at random, the grains have now begun to segregate, with the larger dimer grains moving to the upper layers and the monomers falling into the lower layers.
}
\label{fig:snapshot}
\end{figure}

\hspace{-13pt} half 
the minimum range for interaction between two grains can be defined in 
the case of a strong restoring force. 
If the effective radii of the grains is taken to correspond to an 
experimental radius of 1.65 mm and the gravitational acceleration 
$g = 0.05$ corresponds to 9.8~m/s$^{2}$, we can convert the units of 
the simulation to experimental units.  
The simulation length unit $\Lambda = 0.33 / 0.4$ cm and the time unit 
$\tau = \sqrt{(0.33 \ F_{g})/(r_{g}mg)}$ s.  
Taking the mass of a monomer grain to be 0.2 g, the restoration 
spring constant 
$k_{g} = (20 \ r_{g}mg)/(0.35 \ F_{g})$.  
In this way we can convert the simulated particle velocities to the 
experimental units of cm/s.  

The ratchet sawteeth composing the floor are of a size similar to
that of the grains, as in the experiments of
Ref.~\cite{Farkas99}.
The ratchet profile
(see Fig.~{\ref{fig:snapshot}}) is parameterized in terms of the number of 
sawteeth $N_{s}$, the tooth height $h_{s}$, and the tooth asymmetry $a_{s}$, 
calculated as the width of the left, rising portion of the tooth divided by 
the total width of one tooth.
We take $N_{s} = 16$, $h_{s} = 1.25$ and $a_{s} = 0.0$, 
corresponding to isosceles right triangle-shaped teeth roughly 1.03 cm 
wide and 1.03 cm tall.
This geometry is commonly used in both experiments and 
simulations. 
The angle of the left wall of each tooth is 
$\theta_{L} = \tan^{-1}(h_{s}/(a_{s}\Lambda))$ 
and the angle of the right wall is 
$\theta_{R} = \pi - \tan^{-1}(h_{s}/((1 - a_{s})\Lambda))$, 
where for most simulations the system size $\Lambda = 20$.
Each region bordered by a falling ratchet wall to the left and and a rising 
wall to the right is defined to be one cell.
Although the position of a grain might vary significantly inside one cell, 
this motion averages to zero unless the grain travels to a neighboring cell.

We implement the sawtooth base as a virtual wall composed of mirror 
monomers that provide a normal force away from the wall on each sub-grain 
approaching within a distance  $r_{g}$ of the wall.
Walls composed of uniformly-spaced monomers tend to leak at high driving 
amplitudes and were not used.
One fixed grain is placed at the tip of each sawtooth to prevent the
tip of the sawtooth from lodging between two sub-grains of the longer
grains.

We simulate collections of grains consisting of up to 600 sub-grains 
(corresponding to $N_n =$ 600 monomers, 300 dimers or 200 trimers).  
At the beginning of each simulation the grains are dropped upon a vibrating 
base with an asymmetric sawtooth profile.  In the case of mixtures,

\begin{figure}
 \centerline{
 \epsfxsize=3.3in
 \epsfbox{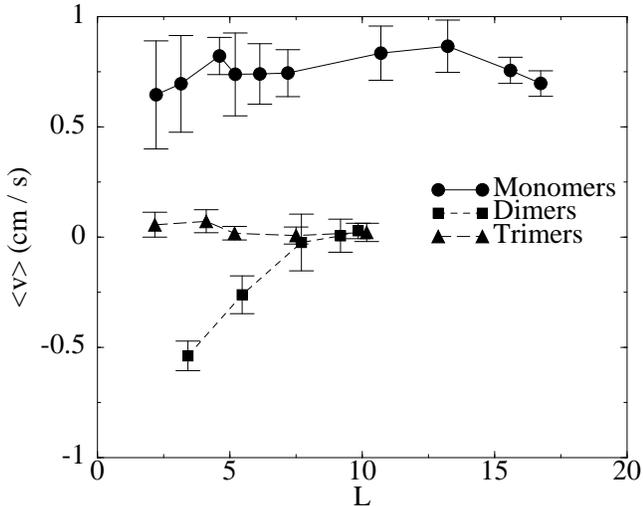}}
\caption{
Average horizontal ratchet velocity $\langle v\rangle$ for monomer, dimer and 
trimer grains as a function of the number of layers of grains $L$.
While monomers move to the right,
dimers move in the {\it opposite} direction.
Trimers are too massive to move at this driving amplitude.
As more and more layers of grains are added, all particles slow down, 
with the heavier particles quenching first.
The downward trend of the monomer velocity continues with increased layers, 
but these points are omitted from the graph due to space.
The sawteeth parameters were $N_{s} = 16$, $h_{s} = 1.25$, and $a_{s} = 0.0$, 
with driving at roughly 69 Hz and $A_{d} = 0.75$.
}
\label{fig:v-layers}
\end{figure}

\hspace{-13pt} 
the order in which grains of different types are dropped
is randomized to avoid artificial stratification.
The standard system used is 200 monomers in a 20 x 20 region with periodic 
boundary conditions in the $x$ (horizontal) direction.  
For a given parameter set the simulated results were divided into ten portions for statistical purposes, consistent with similar techniques applied to experimental data\cite{Farkas99}.
Each simulation is between $10^{6}$ and $10^{7}$ MD steps in length, corresponding to times of roughly $6.5$ to $65$ seconds.

We also performed simulations on systems with $\Lambda=$ 5, 10, and 40 both
to check for finite size effects and
to achieve larger layer thicknesses without increasing the number of 
particles.
Since the ratchet effects observed here occur on length scales equal to the 
size of a single ratchet tooth, varying the size of the system did not 
affect the results obtained.
Reversing the direction of the sawteeth had the expected effect of 
reversing the direction of the grain velocity.

\section{Results}

\subsection{Average horizontal velocity}

We first consider the dependence of the average grain velocity along
the ratchet,
$\langle v\rangle = (1/N_n)
\sum {\bf v}_{i}{\hat x}$, 
on the 

\begin{figure}
\centerline{
\epsfxsize=3.3in
\epsfbox{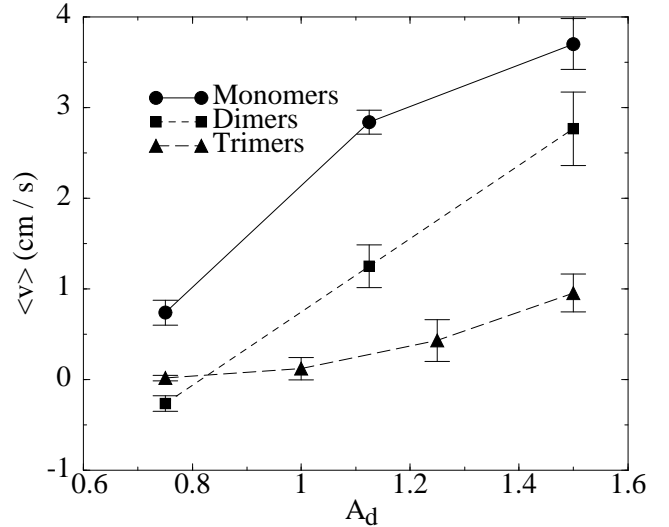}}
\caption{
Average velocity $\langle v\rangle$ as a 
function of driving amplitude $A_{d}$, for 
a sawtooth height $h_{s} = 1.25$.
While the three different types of grains show different behavior 
at the low amplitude of
$A_{d} = 0.75$, 
at high $A_{d}$ all types of grains 
move to the right.
}
\label{fig:v-amplitude}
\end{figure}

\hspace{-13pt}
number of layers $L$ of grains.
To determine $L$, we found the number of layers in the $z$ direction
that were occupied by grains
for more than 80\% of the time, 
and add the 
fraction of time the next highest layer was filled 
\cite{footnote2}.
To measure $\langle v\rangle$, we periodically record the horizontal 
displacement over a given time interval of every particle in the system. 
We also determine $\langle v\rangle$ for particles of a specific 
size, and for particles in a given layer, as well as recording
the average mass and average number of particles in each 
layer.

As shown in Fig.~{\ref{fig:v-layers}},
$\langle v\rangle$ for monomers
initially increases for small numbers of layers, and
gradually decreases to zero for large $L$.
This behavior is in good agreement with
previous experiments \cite{Farkas99} and simulations \cite{simulation1}, and 
is believed to result from three competing effects \cite{simulation1}.
For small $L$ the ratchet velocity increases with $L$ 
because a minimum number of grains is required for every grain to 
fully explore the ratchet geometry and not become trapped.
Because the upper layers tend to move in a direction opposite to that of the 
lower layers, increasing the number of layers eventually decreases the average 
velocity.
Additionally, increasing the number of particles increases the granular 
pressure on the lower layers where the greatest movement normally occurs.

Also shown in Fig.~{\ref{fig:v-layers}} are $\langle v\rangle$
for dimer and trimer grains.
Each type of grain displays a very different response. 
Under conditions where the monomers move to the right, dimers move to the 
{\it left}, while the massive trimers lock together and show very little 
movement.
We have also considered grains composed of four and five subgrains, which produce little or no net movement, similar to the behavior of the trimers.
Simulations of a single 

\begin{figure}
\centerline{
\epsfxsize=3.3in
\epsfbox{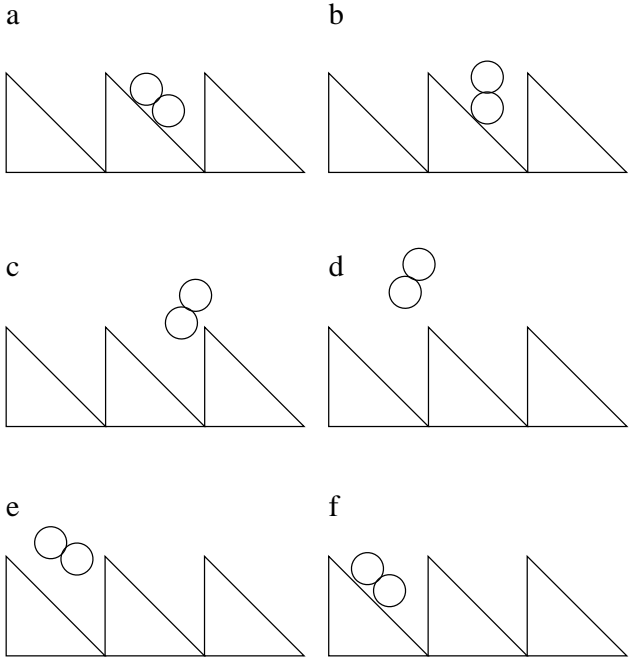}}
\caption{
Illustration of dimer grain motion at consecutive times.
When the dimer strikes the point of a sawtooth at a 
near-perpendicular angle (panel c)
it is propelled backwards into the neighboring cell.
The sloping surface of each tooth tends to push the dimer against the 
rightward point more often than the leftward point, resulting in a net 
velocity to the left.}
\label{fig:dimer-motion}
\end{figure}

\hspace{-13pt}
grain ($N_{n}=1$)
produce ratcheting behavior consistent 
with the value of $\langle v\rangle$ at $L=1$ in Fig.~{\ref{fig:v-layers}}.

The different behaviors of the three grain types is determined both
by grain mass and by grain symmetry.  Longer grains are more massive
and require stronger 
driving to initiate motion.
Figure~{\ref{fig:v-amplitude}} 
illustrates that as the amplitude of the driving 
force is increased, both dimers and trimers can be made to move in the same 
direction as the monomers.
Higher driving amplitudes are also required to maintain the same
amount of motion as the number of layers, and thus the total mass of the
system, increases.

\subsection{Ratchet mechanism}
In the elongated grains, rotation and interactions with
the flat walls of the ratchet play an important role in determining the
ratcheting behavior.
At low driving amplitudes dimers display a ratchet velocity 
$\langle v\rangle$ {\it opposite in 
direction} to the average velocity displayed by monomers under similar 
conditions.  
To explain this, we illustrate the dimer motion in 
Fig.~{\ref{fig:dimer-motion}}.
Dimers are frequently 
flattened against the sloping edge of the ratchet both by pressure from above 
and torque from the moving floor
[Fig.~{\ref{fig:dimer-motion}}(a)]. 
The upward motion of the ratchet then 

\begin{figure}
\centerline{
\epsfxsize=3.3in
\epsfbox{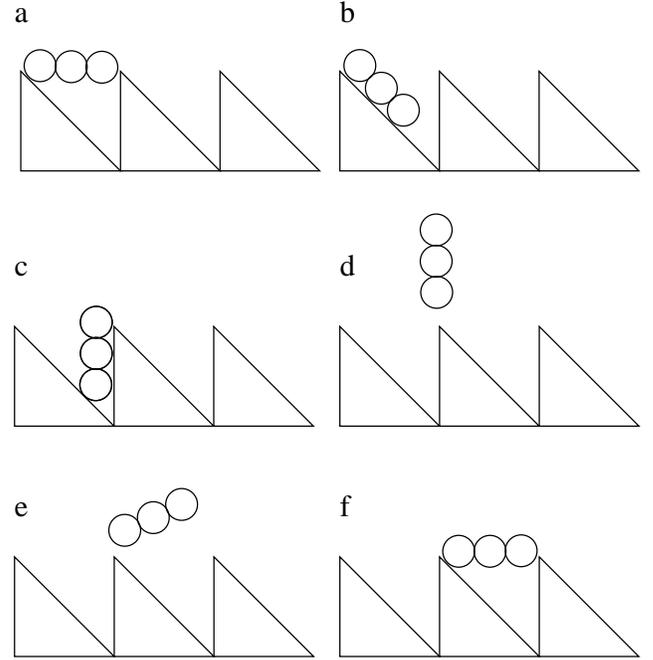}}
\caption{
Illustration of a ratcheting trimer grain at consecutive times.
Trimers tend to be pressed against the vertical edge of the sawteeth, 
where they can then be launched vertically to fall into either the same 
cell or the cell to the right.
Thus, trimers display net motion to the right when driven with
sufficient amplitude.
}
\label{fig:trimer-motion}
\end{figure}

\hspace{-13pt}
launches the dimer into the air
[Fig.~{\ref{fig:dimer-motion}}(b-c)].
If, upon falling, 
the dimer strikes the point of a sawtooth in a glancing blow involving
only one of the two sub-grains, the dimer will be set into a spinning
motion but will gain little translational motion, and will generally
fall back into the same cell or occasionally move one cell to the right.
If instead the midpoint of the dimer collides with 
the sawtooth point,
the dimer receives no torque and does not spin, but is 
kicked a great distance away from the peak
[Fig.~{\ref{fig:dimer-motion}}(c-e)].
These events occur regularly, often kicking the dimer to the neighboring cell 
on either the left or the right, depending on which sawtooth point the 
ratchet encountered with the necessary perpendicular orientation 
[Fig.~{\ref{fig:dimer-motion}}(f)].  Since the sloping surface of each
tooth tends to push the dimer against the rightward point more often
than the leftward point, a net velocity to the {\it left} arises.

At higher driving amplitudes, the dimer is thrown completely over the
sawtooth point into the next cell to the right by the sloping teeth.
This is the mechanism by which the monomers ratchet, and indicates why
at large driving amplitudes all grain types show the same behavior.
It is the interaction of the dimer with the sawtooth point that leads
to the reversed (leftward) velocity.
The significant possibility of being kicked to the right by the sawtooth
point explains 
the relatively small
magnitude of the leftward dimer 
velocity compared to the rightward velocity of the
monomers.
Changing the mass of the sub-grains, and therefore changing the moment of 
inertia of the dimers, reduces or eliminates the net negative velocity, 
indicating the importance of dimer rotation on this effect.

By comparison, Fig.~{\ref{fig:trimer-motion}} shows
the motion of a trimer at a driving amplitude large enough to
produce net grain movement.
For the ratchet arrangement considered here, the 
trimers are large enough to lie across the top of the sawteeth, as in
Fig.~{\ref{fig:trimer-motion}}(a).
Eventually interactions with other trimers and the moving base
work the trimer into the cell [Fig.~{\ref{fig:trimer-motion}}(b)],
where the 
motion of the ratchet swings the trimer until it is aligned vertically, 
flush with the vertical edge of the sawtooth 
[Fig.~{\ref{fig:trimer-motion}}(c)].
The next upward kick from the sawtooth can then propel the trimer above the 
sawteeth
[Fig.~{\ref{fig:trimer-motion}}(d)].
If it moves to the left, the trimer lands in the same cell and is eventually 
aligned and launched again.
If the trimer moves to the right, it lands in the cell neighboring to the 
right and has successfully ratcheted [Fig.~{\ref{fig:trimer-motion}}(e-f)].
Because the trimers are three times as heavy as monomers, a large 
driving amplitude is necessary for this effect to occur.

Simply increasing the width of the sawteeth so that the trimer grains are 
the same size relative to the sawteeth as the dimers under conditions in 
which the dimers moved in the negative direction is not sufficient to make 
the trimer grains move to the left instead of to the right.
Indeed, no change in system geometry produces this effect, since 
the reversed velocity observed for dimers does not occur for trimers due
to the different symmetry of the two grains.  Even if a trimer strikes
the tip of a sawtooth at the center of mass of the trimer, the grain-grain
interactions will shift the sawtooth toward the left or right end of
the trimer.  Here the sawtooth tip will exert a torque on the trimer, causing
the trimer to spin rather than giving the trimer significant translational
motion.  Only interactions with the flat sides of the teeth produce
significant translational motion of the trimers, just as in the case of
monomers.  For longer grains with an even number of sub-grains, 
such as quadrimers, reversed velocity still does not appear.  Here,
it is possible for the sawtooth tip to strike the center of mass of the
quadrimer and launch the grain toward the left without torque.  It
is, however, twice as likely that the sawtooth tip will strike the quadrimer
between an end sub-grain and a central sub-grain, rather than between
the two central sub-grains, and will thus spin the quadrimer rather than
translating it.  In this case, any leftward grain motion is swamped by 
net rightward motion, and no reversed velocity can be observed.  Only
the dimer grains have the proper symmetry to produce leftward velocity.

\subsection{Stratified velocity by layer}
\begin{figure}
\centerline{
\epsfxsize=3.3in
\epsfbox{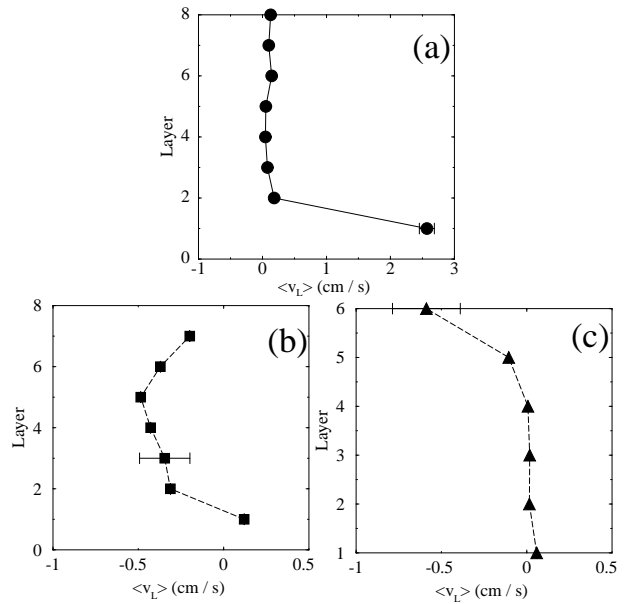}}
\caption{
Stratified velocity $\langle v_{L}\rangle$ as a function of 
height above the base for (a) monomer, (b) dimer, and (c) trimer grains.
The layers responsible for the net motion of the monomers and dimers are 
reversed, with the largest monomer motion occurring close to the ratchet 
floor, while the largest dimer motion occurs well above the ratchet.
Though the top layer of trimer grains moves significantly to the left, this 
layer is very rarely occupied, and represents a negligible contribution to 
the overall velocity.
The lack of variation with height for trimers further indicates that they are 
locked together.
}
\label{fig:v-by-layer}
\end{figure}

When the average velocity is examined layer-by-layer, as 
in Fig.~{\ref{fig:v-by-layer}}, clear differences between monomer, dimer 

\hspace{-13pt}
and 
trimer grains appear.
The layer in which a particular grain is located is determined by counting 
the number of grains in the region bounded vertically by the height of the 
grain and the height of the ratchet below, and horizontally by half the 
length of the grain both to the left and the right.
In this way, the number of layers can be calculated dynamically for grains of 
any size, although extended grains appear to be located higher 
than smaller grains due to the larger amount of space present 
beneath them.
For this reason, when studying mixtures it is necessary to confirm layer 
data by observing animations depicting the actual location and motion of each 
grain.

Fig.~{\ref{fig:v-by-layer}}(a)
shows that the velocity of monomers in the horizontal direction is 
stratified, with the fastest moving layers located nearest to the ratchet 
floor.  This result agrees with previous studies 
of spherical particles \cite{simulation1}, and is consistent with the fact
that grains close to the ratchet teeth have the largest component of their
velocity in the horizontal direction, whereas grains that have been thrown
well above the teeth have most of their velocity in the vertical direction
and show only a small horizontal component of velocity. 
It is interesting to note that the results for horizontal velocity as a 
function of the number of layers at higher amplitudes
is extremely similar to 

\twocolumn[\hsize\textwidth\columnwidth\hsize\csname @twocolumnfalse\endcsname
\begin{table}
\begin{center}
\begin{tabular}{|c|ccc|ccc|} 
\hline
{Amplitude} & 
{$N_{1}$} & 
{$N_{2}$} & 
{$N_{3}$} & 
{$\langle v\rangle_{1}$} & 
{$\langle v\rangle_{2}$} & 
{$\langle v\rangle_{1}$} \\
\hline
0.75 & 200 & - & - & $0.762 \pm 0.109$  cm/s & - & - \\
0.75 & - & 100 & - & - & $-0.263 \pm 0.0693$  cm/s & -\\
0.75 & - & - & 70 & - & - & $0.0173 \pm 0.0306$  cm/s \\
1.5 & - & - & 70 & - & - & $0.955 \pm 0.209$  cm/s \\
0.75 & 100 & 50 & - & 
$0.107 \pm 0.0765$ cm/s &
$-0.123 \pm 0.105$  cm/s & - \\
0.75 & 100 & - & 35 & 
$0.474 \pm 0.0904$ cm/s & - &
$0.134 \pm 0.0645$  cm/s \\
0.75 & - & 50 & 35 & - &
$0.0267 \pm 0.0455$ cm/s &
$0.0274 \pm 0.0446$ cm/s \\
1.5 & - & 50 & 35 & - &
$2.837 \pm 0.179$ cm/s &
$1.018 \pm 0.0765$ cm/s \\
\hline
\end{tabular}
\caption{Ratchet velocity $\langle v\rangle$ for different grain mixtures.  
The driving amplitude is indicated in the left column.
$N_{n}$ is the number of n-mers in a mixture and $\langle v\rangle_{n}$ 
is the corresponding velocity of that grain type.
Each mixture contains approximately 200 sub-grains divided evenly by
mass among the different grain types.  Thus there are fewer dimers and trimers 
than monomers in a given mixture.}
\end{center}
\end{table}
]

\hspace{-13pt}
measurements of 
vertical velocity as a function of layer in confined granular media that 
display size-segregation phenomena when vibrated \cite{JaegerRMP}.

We find a very different behavior for dimers, however.
As shown in Fig.~{\ref{fig:v-by-layer}(b), 
the dimer layers nearest the ratchet 
move slowly, while the {\it higher} layers display the largest velocity. 
This is a result of the dimer transport mechanism, which relies on the
interactions of the dimers with the tips of the ratchet sawteeth.
Dimers lying below the sawteeth tips are trapped in a ratchet cell and,
although they may have significant rotational 
motion, have very little
horizontal velocity.  In contrast those dimers with enough horizontal
velocity to move to the next cell have been launched by the sawtip teeth,
rather than by the floor of the ratchet as in the case of monomers, and
thus move in much higher layers.

The velocity profile of the trimers is flat and close to zero, indicating
that for this driving amplitude the
massive, elongated trimers have locked together.
At higher driving amplitudes when the trimer grains begin to move, the 
velocity profile of 
the trimers is very similar to that of the monomers.

\subsection{Mixtures}
When different types of grains are mixed, we observe size-segregation
phenomena.
Fig.~{\ref{fig:snapshot}} shows a snapshot from a simulation of dimer and 
monomer grains where segregation by size is occurring.  The dimers
rise to the upper layers while the monomers fall toward the bottom layers.
The segregation is independent of the presence of the ratchet floor, as 
it is also observed when the mixture is vibrated on a flat 
surface.
Since the ratchet floor and the periodic boundary conditions in our system 
prevent the formation of convection cells, the segregation observed here 
results when the smaller grains are able to slip around the larger grains and 
move toward the bottom.

The fact that the fastest monomer motion occurs on the lower layers and
the fastest dimer motion occurs on the upper layers, combined with the
fact that dimers tend to sit on the top layers and monomers on the bottom
layers,
suggests that in a mixture the two species could be separated horizontally.
Table I shows average grain velocities by species
for several combinations of grain sizes.
A clear difference in the direction of the grain velocity appears for
the mixture of 100 monomers with an equal mass of 50 dimers.
The monomers move toward the right at $0.107$ cm/s while the dimers
move toward the left at $0.123$ cm/s.
It is important to note that unlike previous work in which segregation 
effects reduce the relative densities of grains in different regions, the 
periodic boundary conditions used here 
keep the relative proportion of monomers to 
dimers constant.
In an actual separation scheme with open boundaries, 
the grains would be able to separate into 
homogeneous groups that could then travel even faster in opposite directions.

As a further probe of the segregation effect,
Fig.~{\ref{fig:mix}} shows the average velocity-by-layer 
$\langle v_{L}\rangle$  and the average mass-by-layer 
$\langle m_{L}\rangle$ for the monomer-dimer mixture.
The segregation by layer of the two grains is apparent in the variation of 
the average mass, and the differences of the average velocities of the layers 
occupied by different grains makes it clear that the two different types of 
particles move in opposite directions.
The decrease in the average mass at the highest layers reflects that these 
layers are not always occupied.

In a mixture of monomers and trimers,
the monomers act as 
{\it lubrication} for the motion of the trimers. 
As indicated in Table I, while the monomer velocity has decreased roughly 
$40\%$ compared to 
a system of pure monomers due to the presence of the trimers,
the trimer velocity has increased nearly {\it ten times} compared to a
system of pure trimers.
The monomers are able to lubricate the system because they are more likely to 
occupy the lower layers where the trimers might otherwise be trapped.
The trimers are then dragged along by the layer of moving monomers beneath 
them.

Results for a mixture of dimers and trimers are presented for both 
the driving at the standard amplitude of $A_d=0.75$ 
and at a doubled amplitude of $A_d=1.5$.
Due to the inverted velocity profile of the dimers relative to the monomers,
the dimers cannot 
lubricate the motion of the trimers in the same way as monomers do.

\begin{figure}
\centerline{\hbox{ \hspace{0.0in} 
    \epsfxsize=1.6in
    \epsffile{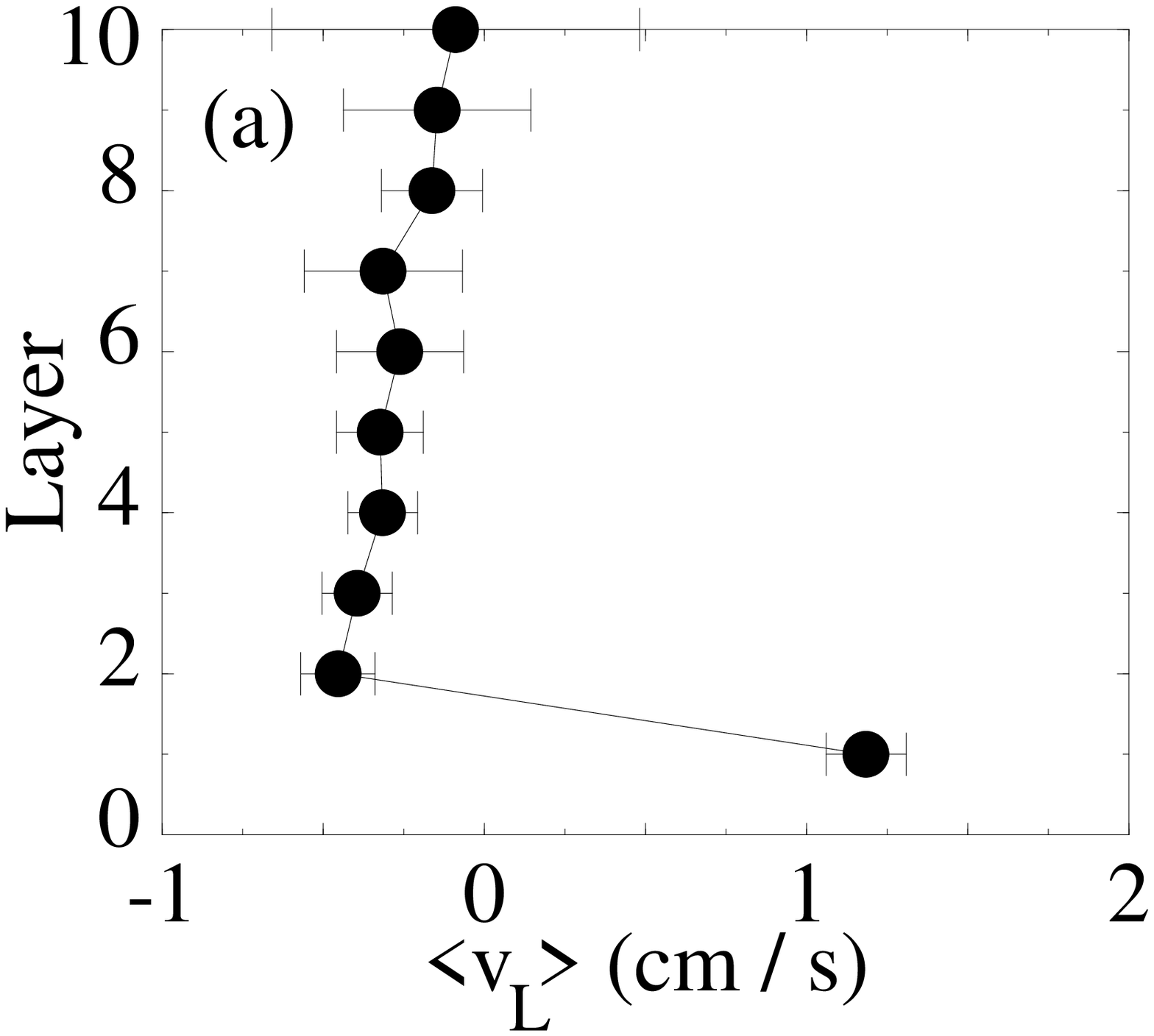}
    \hspace{0.1in}
    \epsfxsize=1.6in
    \epsffile{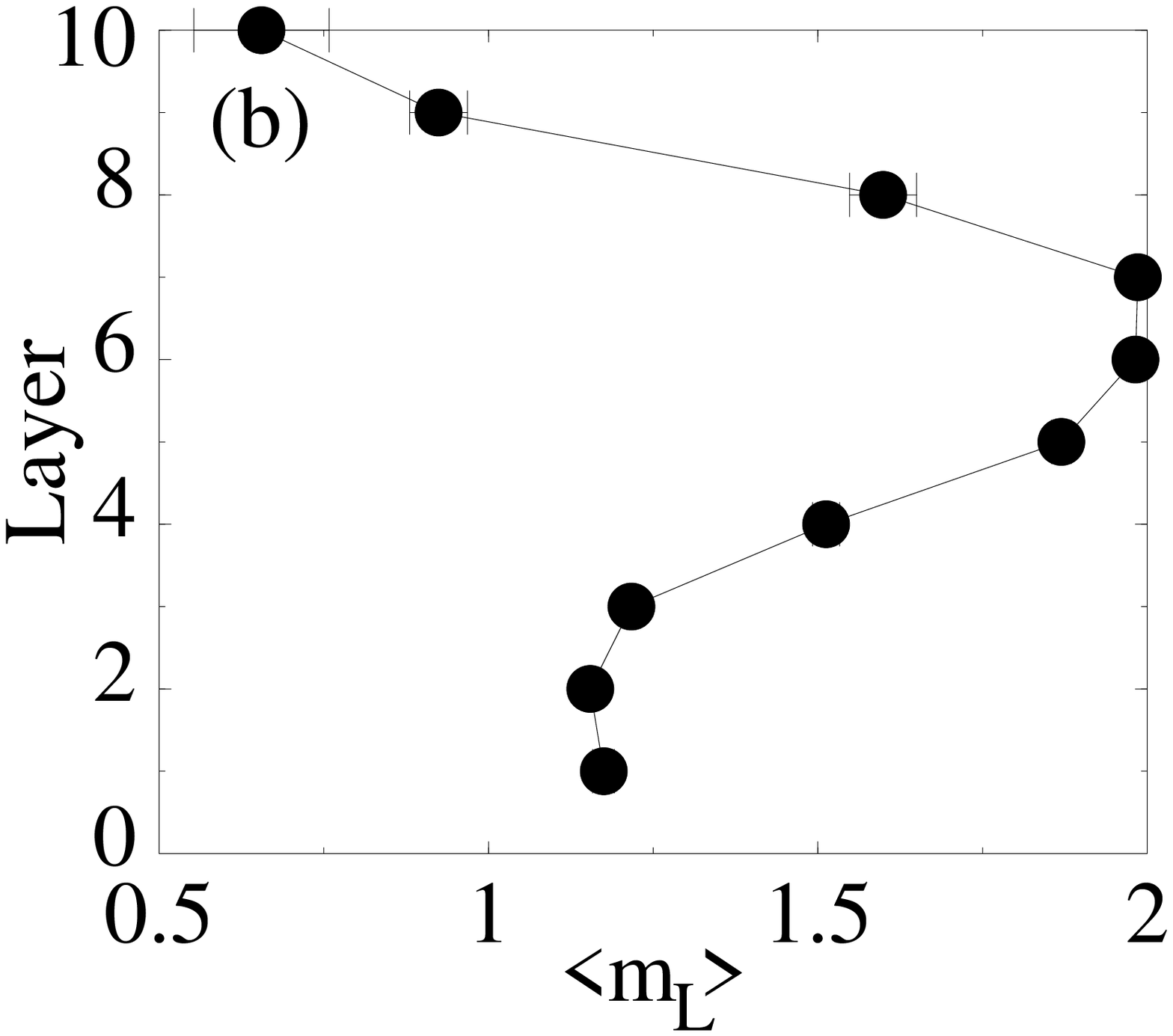}
    }
  }
\caption{
Variation of (a) $\langle v_{L}\rangle$, (b) $\langle m_{L}\rangle$, 
with height above ratchet for a mixture of 100 monomers and 50 dimers.
}
\label{fig:mix}
\end{figure}

\hspace{-13pt}
Because the larger trimers migrate to the higher layers, the dimers are 
trapped in the lower layers where they do not move. 
By filling the lower layers, the dimers prevent the trimers from becoming 
trapped, but there is little horizontal motion propagated through the 
dimers to the trimers to cause movement.
The results for a mixture of 
dimers and trimers at high driving indicates 
that the two types of particles could still be slowly segregated because the 
dimers move three times faster than the trimers 
(as in Fig.~{\ref{fig:v-amplitude}}).

\section{Conclusion}

Using a model for non-spherical granular media, we have considered the
interactions of elongated grains and mixtures of grains with a vibrating
ratchet-shaped base.  Monomers show a net velocity to the right
induced by the asymmetry of the ratchet teeth.  
Dimers have a net velocity to the {\it left} for a range of driving
amplitudes in which they interact with the tips of the sawteeth, due
to the symmetry of the dimer.  
At high driving amplitudes all
grains show monomer-like behavior.  The velocity profile of monomers
consists of fast-moving grains in the bottom layer close to the ratchet
floor, with slow-moving grains on the upper layers.  For dimers, the
velocity profile is inverted and the fastest moving grains are
in the upper layers.  In mixtures of grains, the larger grains move
to the top layers, leading to stratification, and the difference in
velocity of the grain species makes horizontal segregation of the
grain types possible.  Such segregation is especially effective in the
case of mixtures of monomers and dimers, where the two species move in
opposite directions.  Finally, we demonstrated that adding monomers to
a sample of trimers produces a lubrication effect that significantly
increases the velocity of the trimers.

\section*{Acknowledgements}
This work was supported by the US Department of Energy under
contract W-7405-ENG-36.  JFW acknowledges the kind hospitality of CNLS.


\begin{references}
\bibitem{footnote2}For instance, if layers one, two and three were 
occupied more than $95\%$ of the time and layer four was occupied 
$27\%$ of the time, the number of layers would be $3.27$.
\end{references}
\end{document}